\documentclass[9pt,twocolumn,twoside]{osajnl}
\journal{josaa}
\usepackage{tabularx} % for 'tabularx' environment
\usepackage{subcaption}
\usepackage{standalone}
\setboolean{shortarticle}{false}

\title{Impact of time-variant turbulence behavior on prediction for adaptive optics systems}
\author[1,*]{Maaike van Kooten}
\author[1,2]{Niek Doelman}
\author[1]{Matthew Kenworthy}

\affil[1]{Leiden University, Niels Bohrweg 2, Leiden, Netherlands, 2333 CA}
\affil[2]{TNO Technical Sciences, Stieltjesweg 1, 2628 CK Delft, The Netherlands}
\affil[*]{vkooten@strw.leidenuniv.nl} %% email address is required

\begin{abstract}\label{abstract}
For high contrast imaging systems,  the time delay  is one of the major limiting factors for the performance   of the extreme adaptive optics (AO) sub-system and, in turn, the final contrast. The time delay is due to the finite time needed to measure the incoming disturbance and then apply the correction. By predicting the behavior of the atmospheric disturbance over the time delay we can in principle achieve a better AO performance. Atmospheric turbulence parameters which determine the wavefront phase fluctuations have time-varying behavior. We present a stochastic model for wind speed and model time-variant atmospheric turbulence effects using varying wind speed. We test a low-order, data-driven predictor, the linear minimum mean square error predictor, for a near-infrared AO system under varying conditions. Our results show varying wind can have a significant impact on the performance of wavefront prediction, preventing it from reaching optimal performance.  The impact depends on the strength of the wind fluctuations with the greatest loss in expected performance being for high wind speeds.   
\end{abstract}

\setboolean{displaycopyright}{true}

\begin{document}

\maketitle

\section{Introduction}
\label{sec:introduction}
Since the first direct image of an exoplanet~\cite{Chauvin_2004}, high contrast imaging (HCI) has been rapidly advancing. Capable of providing insight into planet formation, characterizing an exoplanet's atmosphere, and searching for signs of habitability, HCI spatially separates the planet's photons from the ones emitted by the host star. With multiple dedicated instruments such as SPHERE on the VLT~\cite{SPHERE}, GPI on Gemini South~\cite{GPI}, and SCExAO on Subaru~\cite{SCEXAO} many new observing techniques, better coronagraphs, and low noise detectors have been developed and tested. These efforts have resulted in processed contrast levels of $10^{-4} - 10^{-5}$ at $0.5"-1"$ separation in the near infrared\cite{Otten_2017}. To find cold jupiters, neptunes, and rocky planets,  we need improve the contrast levels for HCI systems; aiming for $10^{-7}-10^{-9}$ at separations of $0.05"-1"$ at visible wavelengths. 

Currently, one of the most significant limiting factors of a HCI system is the performance of the adaptive optics (AO) sub-system\cite{Kasper_2012}. Specifically, the time delay (which leads to the servo-lag error) prevents us from observing at small separations as a halo of speckles builds up close to the host star during a science exposure \cite{Milli_2017}. In order to improve the performance of the HCI system, the servo-lag error needs to be reduced~\cite{Kasper_2012}. One approach to do this is to run the AO system faster, requiring higher speed deformable mirrors, fast wavefront sensors, and bright guide stars,  thereby reducing the length of the lag  . Alternatively, one can focus on understanding how the phase evolves during the delay and subsequently account for the dynamic behavior during the servo-lag. This is done through predictive control methods, where recent and/or nearby wavefront sensor measurements are used to predict the disturbance over the time lag.

 \subsection{Predictive control in adaptive optics}
 \label{sec:predictivecontrol_AO_intro}
Predictive control to mitigate the effect of time delays in AO systems has been worked on for many years.   Outside of HCI, predictive control has been studied to improve the performance of flux-limited laser guide star AO, increase sky coverage for natural guide stars ~\cite{Jackson_15}, as well as many other AO applications.    Linear prediction has been very successful for tip-tilt compensation, especially when induced by structural vibrations~\cite{Petit_2008}, using Linear Quadratic Gaussian control (LQG). Work on the LQG controller for full atmospheric disturbance compensation has also been done~\cite{Roux_2004}. Different predictors, closely related to the Kalman filter, have allowed for filtering in real time and are now being used in laboratory testing as well as in on-sky testing~\cite{Paschall_1993, Roux_2004, Looze_2009,Gray_2014, Massioni_2011,Correia_2015}. The closely related data-driven H2 optimal controller \cite{Doelman_2009,Hinnen_07, Doelman2_2011} was also tested on-sky,  for tip/tilt modes,            showing a reduction in the temporal error. More recently, the Empirical Orthogonal Functions as a predictor has been shown in theory to improve the contrast of a HCI system by a couple orders of magnitude \cite{Guyon_2017}.  Many of these approaches are based on the assumption that atmospheric turbulence induced wavefront phase fluctuations are a stationary stochastic process. However,  as the turbulence parameters have time-varying behavior, the mean and variance of the wavefront phase are also time-varying, leading to a time-variant process.  Adaptive predictors have been proposed to track the time-varying behavior of statistics of atmospheric turbulence \cite{Maaike_2017}. Although some predictors might be robust against time-varying behavior, none have been rigorously tested.  Therefore, in this work, we look at basic data-driven prediction and test different implementations under time-varying conditions.   

\subsection{Time-variant Atmospheric Turbulence }
\label{sec:TV_turbulence}
Predictive controllers inherently depend on knowledge of turbulence parameters, as well as the dynamical behavior of the AO system. We know the relevant turbulence parameters such as coherence length, outerscale, and wind speed vary, resulting in variations of the mean and variance of phase aberrations. Atmospheric turbulence induced phase fluctuations are essentially time-variant. If we account for this behavior in simulation, we will be able to determine the behavior of a predictor under time-varying conditions.

\subsection{Simulating atmospheric phase}
\label{sec:Sim_phase}
The most recent algorithms for generating atmospheric phase disturbance in simulation have been proposed to minimize computation time ~\cite{Assemat_2006} as well as include the effects of boiling~\cite{Srinath_2015}. They also have flexible architectures that allow for the wind speed, outerscale and Fried parameter to vary at each time step. However, they do not propose how these parameters should vary in time. Without information on how these parameters vary, we are unable to test new control techniques in a simulation environment that reflects all the behavior of the disturbance seen on-sky and instead can only test them under the average behavior. 

 \subsection{Outline}
 \label{sec:outline}
In this paper, we focus on understanding the atmospheric wavefront phase fluctuations of an AO system on short time scales and how the disturbance's time-varying behavior affects a data-driven prediction algorithm. In Sect.~\ref{sec:disturbance_model}, we look at the dynamical behavior of relevant turbulence parameters - including the wind vector, outerscale, and the Fried parameter - used to describe atmospheric wavefront phase fluctuations. We then  present our simulation environment in Sect.~\ref{sec:methods} and introduce a data-driven prediction algorithm in Sect.~\ref{sec:prediction}. We present our results for varying wind in Sect.~\ref{sec:results} and discuss their impact in Sect.~\ref{sec:discussion}. Our future work is proposed in Sect.~\ref{sec:conclusion}.
%%%%%%%%%%%%%%%%%%%%%%%%%%%%%%%%%%%%%%%%%%%%%%%%%%%%%%%%%%%%%%%%%%%%
\section{Describing wavefront phase fluctuations} \label{sec:disturbance_model}
In AO the atmospheric wavefront phase fluctuations are often described as a stationary process with a given variance represented by a spatial covariance function. By assuming Taylor's Frozen Flow hypothesis, we can extend this spatial covariance function to a temporal covariance function through relation with a constant wind speed. Within this framework we can test many different   atmospheric turbulence conditions. 

Extending the above to time variant wavefront phase fluctuations, we vary the wind speed ($v$), the Fried parameter ($r_0$), and outerscale ($L_0$). 
%\textbf{List of what I want in this section}
%\begin{enumerate}
%\item Assemat spatial and temporal covariance function
%\item What is non-stationarity
%\item Show I have a model that is non stationary --PSD at each timestep so that you get a two dimensional image that has time on x axis, frequency on y axis and the intensity is the PSD. Then we can show how the model evolves in time. 
%\item How do we think wind varies, fried parameter varies, and out scale varies
%\end{enumerate}
%%%%%%%%%%%%%%%%%%%%%%%%%%%%%%%%%%%%%%%%%%%%%%%%%%%%%%%%%%%%%%%%%%%%%%%%%%%%%
\subsection{Atmospheric Covariance function} \label{sec:covariance}
We can describe the spatial covariance of  atmospheric phase fluctuations, $\alpha$,  by the von Karman model~\cite{conan_thesis} as in Eq.~\ref{eq:spatial_cov}. 

\begin{multline}
\label{eq:spatial_cov}
 C_{\alpha } (r)=\left( \frac{L_0}{r_0} \right) ^{5/3} \frac{\Gamma(11/6)}{2^{5/6}\pi^{8/3}} \left( \frac{24 \Gamma(6/5)}{5} \right) ^{5/6} \left(\frac{2\pi r }{L_0}\right)^{5/6} \\ \cdot K_{5/6}({2\pi r }/{L_0})
\end{multline}

where $\Gamma (x)$ is the gamma function and $K_u(x)$ is the modified Bessel function of order $u$.

Using the Frozen Flow hypothesis, we can describe a spatial separation, $r$, as a shift over a short period of time, $t$, due to $v$. The temporal covariance follows when this is applied to Eq.~\ref{eq:spatial_cov}.

The wind speed then influences the temporal covariance function. The spatial and temporal covariances are both influenced by $L_0$ and $r_0$. We therefore turn our attention to understanding how these three parameters change in time. First we present a model for the wind vector, then we briefly discuss the behavior of the $r_0$ and $L_0$ as they are understood in the above covariance functions.  

\subsubsection{Wind Vector} \label{sec:wind}
Taking wind data from the Thirty Metre Telescope (TMT) site testing campaign at Mauna Kea~\cite{Schock_2009} we look at the temporal behavior of the wind speed. The data is from a sonic anemometer at a height of 7~m above the summit of Mauna Kea, with a variable sampling frequency between 10-60 Hz. We look at the wind speed and direction fluctuations over a period of three years (2006-2008). 

\begin{figure}
\centering
\includegraphics[origin=c, trim={0cm 1.2cm 0cm 0.2cm},clip, width=\linewidth]{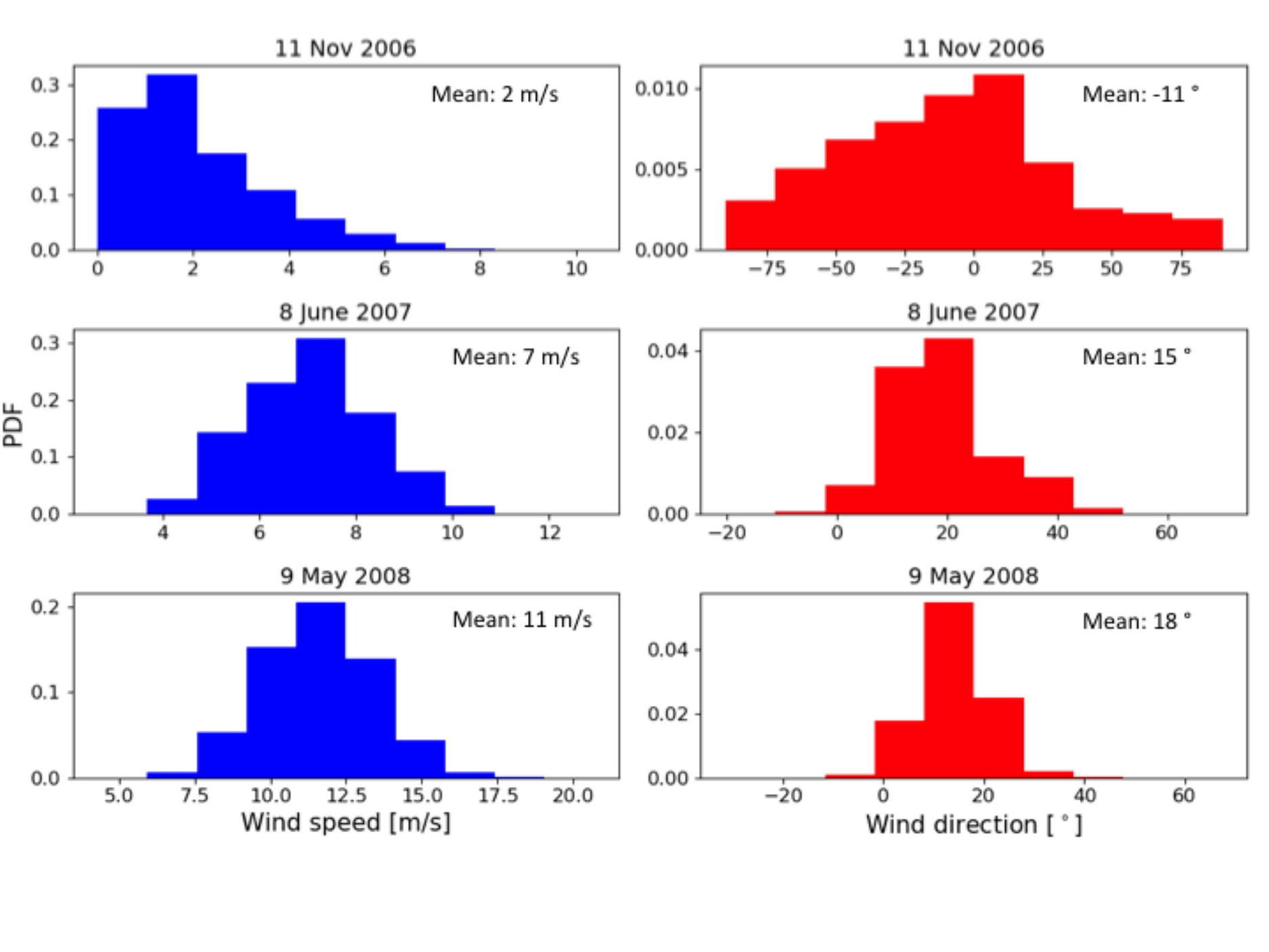}
\caption{Histograms of wind direction and wind speed for three different nights that correspond to the cases in Tab.~\ref{table:1}.}
\label{fig:three_wind_histogram}
\end{figure}

\begin{figure}
\centering
\includegraphics[trim={0cm 0cm 0cm 0cm},clip,width=\linewidth]{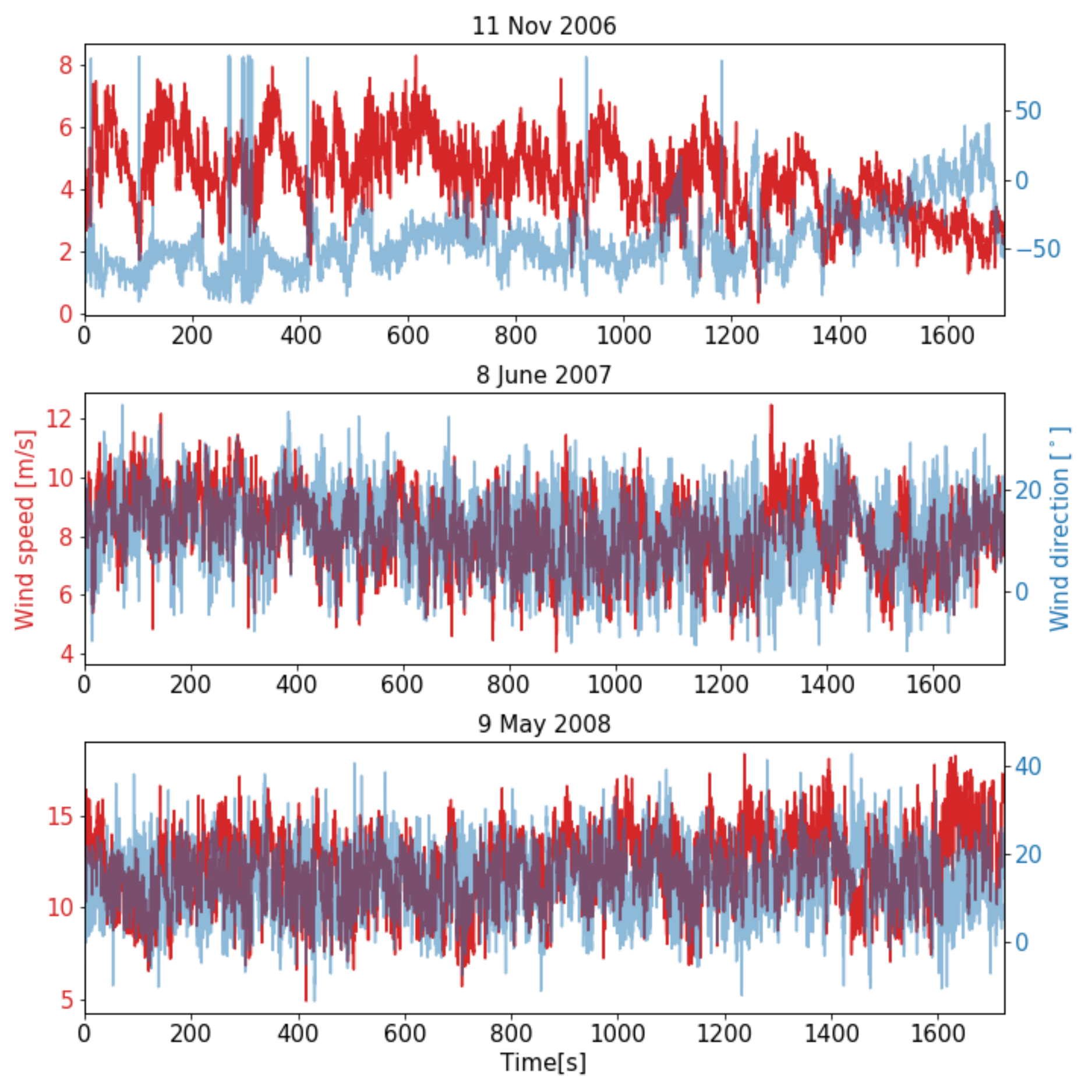}
\caption{ Raw  time series for three different nights showing the classification in Tab.~\ref{table:1}.  The data is first filtered using a Butterworth filter to remove a measurement artifact at 20 Hz due to the sonic anemometer electronics before feeding into our simulations. }
\label{fig:timeseries_wind}
\end{figure}

From this data set we are able to define a stochastic model that describes the averaged power spectral density (PSD) from night time wind speed data for the ground layer. For each night we look at wind speeds faster than 1~m/s. We observe substantial fluctuations in wind speed (greater than 1~m/s) and wind direction (tens of degrees) at frequencies of 1~Hz or greater. We observe 3 types of wind behavior from our analysis that represent a large portion of the data (Fig.~\ref{fig:timeseries_wind}).  In general, over the course of the night either the wind direction or the wind speed can be the dominate source for the fluctuating wind vector. We do not observe nights where both change by large amounts at the same time. Besides these two cases we also have nights where the mean wind speed and wind direction as well as their variances stay the same throughout the night. We consider this to be a realistic Frozen Flow case. These 3 wind behaviors observed is summarized in Tab.~\ref{table:1}. We chose to ignore wind direction the remainder of this work due to large fluctuations in direction for slow wind speeds~\cite{Maaike_2018}.

\begin{figure}
\centering
\includegraphics[trim={0 6cm 0 7cm},clip,width=\linewidth]{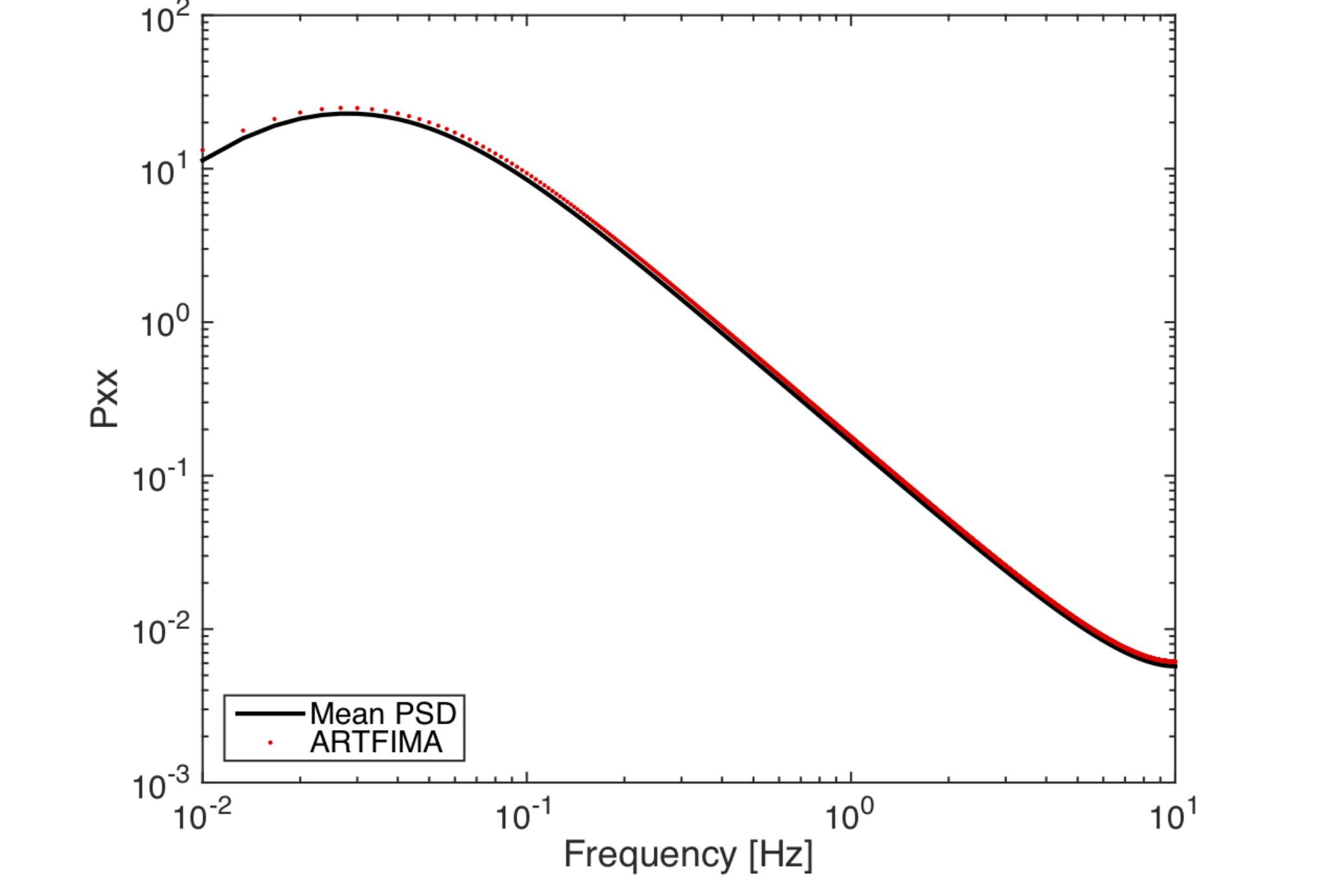}
\caption{Fit of the ARTFIMA model. The black line shows the averaged PSD determined from the data while the red shows the PSD from the ARTFIMA model. }
\label{fig:wind_psd}
\end{figure}

\begin{table*}
\centering
\begin{tabularx}{\textwidth}{XXXX}
 \hline
Classification & Wind Speed & Wind direction & Example nights \\ 
 \hline\hline
Varying wind direction & Slow (<5 m/s) & Large variance (+/- 100 /degrees) & 2006.11.22\\ 
Realistic Frozen Flow & Fast (>8 m/s) with medium variance & Small variance &  2008.05.09 \\
Varying wind speed & Large variance (0-20m/s) & Small variance & 2007.06.15 \\
 \hline
\end{tabularx}
\caption{Three classifications of wind vector behavior seen in our data.}
\label{table:1}
\end{table*}
To estimate a wind speed PSD we split the night into shorter time segments of 10 minutes. We estimate the PSD using the Lomb-Scargle approach for each segment after de-trending and then take the average to determine the overall PSD for that night. We do this for all nights available, determining an average PSD for the wind speed during the period of observations, Fig.~\ref{fig:wind_psd}. A more in-depth study of the wind statistics and how we estimated the wind's power spectral density is in van Kooten et al.~\cite{Maaike_2018}. Here we extend the stochastic model we previously found to capture the roll-off at low frequencies by extending the order of the model proposed to include a tempering component. The resulting stochastic model is a tempered fractionally integrated autoregressive moving average model (ARTFIMA)~\cite{Meerschaert_2013}. Eq.~\ref{eq:fit} shows the PSD for our model. 

\begin{equation}
\Phi (k)=|\frac{(1-1.0047e^{-ik}) }{(1-0.999e^{-ik}) }| ^{2}  |(1-e^{-0.016-ik}) | ^{-2(0.91)}
\label{eq:fit}
\end{equation}
with the angular frequency $k=2\pi~f/f_s$ with $f$ the frequency ranging from 0.002~Hz to 10~Hz and $f_s$=20~Hz.

We present this model   for ground layer wind speed on Mauna Kea for the TMT site location to be used to generate time variant turbulence simulations via varying wind. This model provides a good fit to the individual nightly PSD as well as the averaged PSDs, as the overall shape of the PSDs remains constant. The gain of the nightly PSD does vary by one order of magnitude throughout the data set.

\subsubsection{Fried Parameter}\label{sec:fried}
The Fried parameter ($r_0$) is considered the main turbulence parameter in astronomy that directly influences the achievable imaging resolution.  In reality, $r_0$ varies in time with the $C_n^2$ profile in the atmosphere.   

A dynamic $r_0$ model~\cite{Doelman_2009} was determined from 18 nights of data from La Palma. The model was estimated from data with sampling times ranging from 80s to 120s. This model, therefore, does not include how $r_0$ behaviors at sub-second timescales and we do not know how it varies on AO timescales,    as no other data is available. Therefore, we do not include time-varying $r_0$ in our simulations. 

%%%%%%%%%%%%%%%%%%%%%%%%%%%%%%%%%%%%%%%%%%%%%%%%%%%%%%%%%%%%%%%%%%%%%%%%%%%%
\subsubsection{Outerscale} \label{sec:outerscaletheory}
Many efforts have focused on accurately determining $L_0$ of turbulence with estimations ranging from 5-300~m \cite{Ziad_2016}. From observational campaigns, average values, probability density functions (PDF), and even a time series over the course of a night have been reported. However, the insights only apply to longer timescales, meaning we do not have sufficient knowledge to model the behavior of $L_0$ for AO control loop timescales. 

%%%%%%%%%%%%%%%%%%%%%%%%%%%%%%%%%%%%%%%%%%%%%%%%%%%
\section{Methods} \label{sec:methods}
%\textbf{List of what I want in this section}
%\begin{enumerate}
%\item Explain the 3 different LMMSE filters (recursive, steady state, and adaptive) - model independent
%\item What PSD we use and how do we find the prediction matrix (theoretical representation)?
%\item Show that the design of the controllers/predictors are dependant on windspeed, outer scale, and Fried parameter so that we can flow easily into the next section. 
%\end{enumerate}
 
%%%%%%%%%%%%%%%%%%%%%%%%%%%%%%%%%%%%%%%%%%%%%%%%%%%%%%%%%%%%%%%%%%%%%%%%%%%%%
\subsection{Turbulence simulation}\label{sec:tur_simulation}
We use the method developed by Assemat et al.~\cite{Assemat_2006} to generate  phase screens with von Karman statistics. It has been included in the HCI Python package, HCIPy~\cite{emiel_2018} being developed at Leiden Observatory. At each time step, the phase screen is updated by the algorithm, allowing for infinite phase screen generation as well as continuous updating of the wind vector (speed and direction), $r_0$, and $L_0$. This framework allows for time-variant behavior to be simulated in an efficient way. 

We chose our AO system configuration to be representative of an Extremely Large Telescope (ELT) scale system for HCI based on the planned METIS system for the ELT~\cite{METIS}. We model a single layer atmosphere. See Tab.~\ref{table:config} for configuration details. We assume a idealized wavefront sensor  where each sub-aperture measures a phase point of the wavefront at a given point in time and space.  The total additional noise not due to the servo-lag error has a variance 50 times smaller than the full wavefront phase variance. These choices allow us to focus on the effects of the delay and how a predictor performs. 

\begin{table*}
\centering
\begin{tabularx}{\textwidth}{XXXXXXXX}
\hline
$\lambda _{WFS}$ & D [m]& $N_{WFS}$ & S.A. size [m] & $r_0$ [m] & Frame delay & Loop speed \\ 
 \hline\hline
 2.2 $\mu$ m  & 39 & $74 ^2$ & 0.5 & 0.77 & 2 samples = 2 ms & 1 kHz\\ 
\end{tabularx}
\caption{Simulation parameters for our AO system including the sub-aperture (S.A.) size.}
\label{table:config}
\end{table*}

For the remainder of this work we assume a constant Fried parameter of 13~cm at 500~nm. For stationary wind cases we use a constant wind speed of 12 m/s unless indicated otherwise. Finally, we assume a constant $L_0$ of 24~m unless indicated otherwise.  

In our simulations,  we predict using the open-loop phase and estimate the error at each time step for our predictor. An open-loop or closed-loop setting would make no difference with respect to the predictor performance only. In practice , the deformable mirror and wavefront sensor response can affect the residual wavefront error when predicting in closed-loop.  This must be considered in the AO control design.  

\subsubsection{Wind}\label{sec:wind_simulation}
Since we have real wind time series available, we use this in our simulations. From the site testing data from TMT, we classify the behavior of the wind vector into three classes as shown in Tab. \ref{table:1}. Six 5~s sections (separated by at least 10 minutes in time) for each case are chosen for three days that are representative cases of our classification. The data are linearly interpolated to the AO loop speed allowing the wind speed to be updated at each time step in the simulation. 

%%%%%%%%%%%%%%%%%%%%%%%%%%%%%%%%%%%%%%%%%%%%%%%%%%%%%%%%%%%%%%%%%%%%%%%%%%%%%
\subsection{Data-driven prediction}\label{sec:prediction}
In section~\ref{sec:disturbance_model} the time-varying behavior of $v$, $r_0$, and $L_0$ are described and well as the difficulties in developing a time variant model for atmospheric turbulence. Currently, we only have a model for ground layer $v$ which is site dependent. We, therefore, turn to data-driven methods which use on-sky data to determine the predictor; specifically the linear minimum mean square error (LMMSE) predictor.  In this section, we present the framework for the LMMSE predictor and in Sect.~\ref{sec:compare_predictors} we place our work in relation to other prediction methods for AO.  

We denote the phase  at a single point $i$ of a phase screen at time t, $y_i (t)$, and $\vec{u(t)}$ a P$\times$1 column vector containing a collection of $P$ phase values on a discrete spatial grid at time $t$ (therefore $P$ is the spatial order). We then assume that the future value of a given phase point, $\hat{y}_i$ at the discrete time index $t+d$, is a linear combination of the most recent phase values (our regressors) at time $t$. Note that we do not have to use all the phase values on our spatial grid but are able to select the spatial order by choosing a region around our point of interest (changing P). We use $\vec{a}_i$ to represent the predictor coefficients. With the above, we define the following where we use $T$ to denote the transpose:

\begin{equation}
\hat{y}_i(t+d)=\vec{a}_i ^{T}\cdot \vec{u}(t)
\label{eq:pred}
\end{equation}

\begin{equation}
\vec{u}(t)=
\begin{pmatrix}
y_0(t)  & y_1(t) & y_2(t)& ...& y_P(t) 
\end{pmatrix}^{T}
\end{equation}

We can expand this to include a set of Q most recent measurements and thereby including more temporal information.

\begin{equation}
\vec{w}=
 \begin{pmatrix}
  \vec{u}(t)^{T}  & \vec{u}(t-1)^{T} & \vec{u}(t-2 )^{T} & ...& \vec{u}(t-Q)^{T} 
 \end{pmatrix}^{T}
  \label{eq:w}
\end{equation}
 
We therefore allow for both spatial and temporal regressors through Eq.~\ref{eq:w}, where $w$ is a vector of $PQ\times$1. We have a spatial order of P and a temporal order of Q resulting in the total of $PQ$ regressors. 

With $ < > _t$ being the time average operator, we minimize the cost function:
\begin{equation}
min_{\vec{a}_{i}} <||{y}_i(t+d)-\vec{a}_i ^{T} \vec{w}(t)|| ^2 >_t
\label{eq:mmse}
\end{equation}
finding the vector of predictor (filter) coefficients, $\vec{a}_i$, now a $PQ\times$1 vector. Eq.~\ref{eq:mmse} represents the time-averaged mean square prediction error. 

The strength of this method is that the algorithm immediately extracts any spatio-temporal information from the data - the regressors - provided. It is important to realize for AO that the temporal sampling is much finer than the spatial sampling.
%%%%%%%%%%%%%%%%%%%%%%%%%%%%%%%%%%%
%%%%%%%%%%%% J Comment %%%%%%%%%%%%
%%%%%%%%%%%%%%%%%%%%%%%%%%%%%%%%%%%
%  The above is an interesting statement. I know what you mean, but comparing temporal to spatial sampling is a bit ``apples to pears'' as they have different dimensions. Perhaps say that it is with respect to the spatio-temporal properties of the turbulence itself? Or that it relates to how much data you have available? 
%%%%%%%%%%%%%%%%%%%%%%%%%%%%%%%%%%%
%%%%%%%%%%%%%%%%%%%%%%%%%%%%%%%%%%%
%%%%%%%%%%%%%%%%%%%%%%%%%%%%%%%%%%%
Solving Eq.~\ref{eq:mmse} for our zero-mean stochastic process, the solution can be written in terms of the inverse of the auto-covariance matrix and cross-covariance vector~\cite{Haykin_2002}.
\begin{equation}
\vec{a}_i=\mathbf{C}_{\vec{w}\vec{w}}^{+}\vec{c}_{\vec{w}{y}_i}
\label{eq:recursive}
\end{equation}
where $+$ denotes a pseudo inverse. $\mathbf{C}_{\vec{w}\vec{w}}$ is the auto-covariance matrix of $\vec{w}$, the vector containing the regressors, and ${c}_{\vec{w}{y}_i}$ is the vector containing the cross-correlation between the true phase value, ${y}_i$ and $\vec{w}$. 

Looking at Eq.~\ref{eq:recursive} the optimal AO predictor based on the spatial covariance (Eq.~\ref{eq:spatial_cov}) can be derived for stationary turbulence~\cite{Beghi_2007}. However, we choose a data-driven approach where we estimate the covariances using measured data collected over a short period. During the collection period (or training period), this batch LMMSE predictor cannot be used (or we can use a previously calculated solution) and is considered off-line.

In the batch-wise approach, once the prediction vector is found it is fixed for the rest of the simulation. By making use of the matrix inversion lemma, we can form a recursive solution in which    the pseudo-inverse of the auto-covariance is updated according to Eq.~\ref{eq:cov1}. 

\begin{equation}
\mathbf{C}_{\vec{w}\vec{w}}^{+}(t-d)=\mathbf{C}_{\vec{w}\vec{w}}^{+}(t-d-1)-
\vec{k}(t-d)\vec{w}^T(t-d)\mathbf{C}_{\vec{w}\vec{w}}^{+}(t-d-1)
\label{eq:cov1}
\end{equation}
with
\begin{equation}
\vec{k}(t-d)=\frac{\mathbf{C}_{\vec{w}\vec{w}}^{+}(t-d-1) \vec{w}(t-d)}{1+\vec{w}^T(t-d)\mathbf{C}_{\vec{w}\vec{w}}^{+}(t-d-1)\vec{w}(t-d)}
\label{eq:cov2}
\end{equation}

We can then update the prediction coefficients, $\vec{a}_i(t)$:
\begin{equation}
\vec{a}_i (t)=\vec{a}_i(t-1)+\vec{k}(t-d)\left(y_i(t)-\vec{a}_i(t-1)^T \vec{w}(t-d)\right)
\label{eq:a_update}
\end{equation}

The recursive solution goes on-line immediately with the initial covariance being set to diagonal matrices with large values (as done with recursive least-squares methods). The solution is quick to converge compared to gradient-based methods. The recursive solution depends on all previous measurements finding the optimal solution for all previous data. 

\subsection{The effect of atmospheric parameters on prediction}\label{sec:effect_parameter_prediction}
As mentioned above, using the covariance in Sect.~\ref{sec:covariance} we can derive our optimal predictor in terms of the theoretical covariance. We can then see how the predictor depends on the atmospheric parameters.

For example, the diagonal elements in $\mathbf{C}_{\vec{w}\vec{w}}$~\cite{conan_thesis} are given by:
\begin{equation}
\label{eq:spatial_auto}
 C_{\alpha } (0)= \frac{L_0}{r_0} ^{5/3} \frac{\Gamma(11/6)}{2 \pi^{8/3}}\left(\frac{24\Gamma(6/5)}{5}\right)^{5/6}
\end{equation}
while the off-diagonal elements are given by Eq.~\ref{eq:spatial_cov}. We can then represent $\mathbf{C}_{\vec{w}\vec{w}}$ as a matrix $\mathbf{F}_{\vec{w}\vec{w}}$ multiplied by the scalar $ C_{\alpha } (0)$. Therefore: 

\begin{equation}
\mathbf{C}_{\vec{w}\vec{w}} ^+ =\frac{1}{ C_{\alpha } (0)} \mathbf{F}_{\vec{w}\vec{w}}^+
\label{eq:cww}
\end{equation}

Similarly, we can represent the cross-covariance vector as:
\begin{equation}
\vec{c}_{\vec{w}{y}_i}= C_{\alpha } (0){f}_{\vec{w}y_i}
\label{eq:cwy}
\end{equation}

Inserting Eqs.~\ref{eq:cwy} and~\ref{eq:cww} into Eq~\ref{eq:recursive} we obtain the following: 
\begin{equation}
\label{eq:theoretical_pred}
\vec{a}_i=\mathbf{F}_{\vec{w}\vec{w}}^+ {f}_{\vec{w}y_i}
\end{equation}
 
Eq.~\ref{eq:theoretical_pred} does not depend on $r_0$ (only present in $C_{\alpha } (0)$), but is instead dependent on $L_0$ and $v$ (applying Frozen Flow approximation). Therefore, when $v$ or $L_0$ vary our optimal predictor changes as well.  

\subsection{Dealing with time-varying behavior}\label{TV_behaviour}
Both the batch and recursive methods are originally designed for stationary problems and so we apply them to our stationary case. To make these methods suitable for time varying cases, the batch can be retrained and reset as frequently as needed. We can introduce a forgetting factor for the recursive LMMSE that is applied during the predictor update, Eq.~\ref{eq:cov1}. This allows for the recursive method to ignore the previous states of the turbulence by gradually reducing the weighting of old observations in the prediction. 

\subsection{Comparison to other predictors}\label{sec:compare_predictors}
Before looking at the performance and behavior of the LMMSE predictor we first put our work in context of prediction for AO. We aim to give    a brief overview of most prediction methods in AO with particular attention to the predictor structure.  The batch LMMSE is very similar to the Empirical Orthogonal Function~\cite{Guyon_2017} approach with a slightly different implementation for the matrix inversion. Many other data-driven predictors have been proposed including the traveling wave predictor~\cite{Hardy_1998} that makes use of the Frozen Flow hypothesis. By knowing the wind vector, one can spatially shift the current phase measurements; predicting what the wavefront will look like at the time of our correction. We can therefore represent the traveling wave predictor in our own framework where we do not need knowledge of the wind speed but instead it is extracted intrinsically in the LMMSE algorithm. More specifically, if we choose our regressors such that we only allow the most recent measurement, our LMMSE will find a pure spatial solution that is equivalent to the traveling wave predictor. Finally, we can have a pure temporal solution that only uses recent measurements for the phase point of interest. When we limit ourselves to the most recent measurement only for the batch-wise LMMSE predictor, we get the equivalent of the closed-loop integrator with some gain. The integrator and our s1t1 are often not classified as a predictor; however, they are sometimes considered zero order predictors~\cite{Hardy_1998}. If we include multiple previous measurements, we can form a higher order temporal predictor that is similar to an AR-structure approach \cite{Roux_2004} (that might be used in the optimal control framework).  
 Within the multi-congugate AO (MCAO) community, the inclusion of a prediction step has result in the spatial angular (SA) predictor \cite{Correia_2015, Jackson_15} as well as a moving grid prediction included in the DM fitting step ~\cite{Piatrou_07}. The essence of these approaches are similar to the LMMSE, making use of either temporal or spatial regressors.  

%%%%%%%%%%%%%%%%%%%%%%%%%%%%%%%%%%%%%%%%%%%%%%%%%%%%%%%%%%%%%%%%%%%%%%%%%%%%%
\section{Simulation Experiments}\label{sec:results}
In this section, we present and discuss a number of different simulation results, starting with the case of stationary turbulence in order to gain intuition on how the LMMSE behaves under these conditions. We then present the results for time variant turbulence, applying the LMMSE predictor for the case of varying wind speed from field measurements.

\begin{figure*} % "[t!]" placement specifier just for this example
\begin{subfigure}{0.23\textwidth}
\includegraphics[height=7cm, trim={2cm 0 0 0},clip,width=\linewidth]{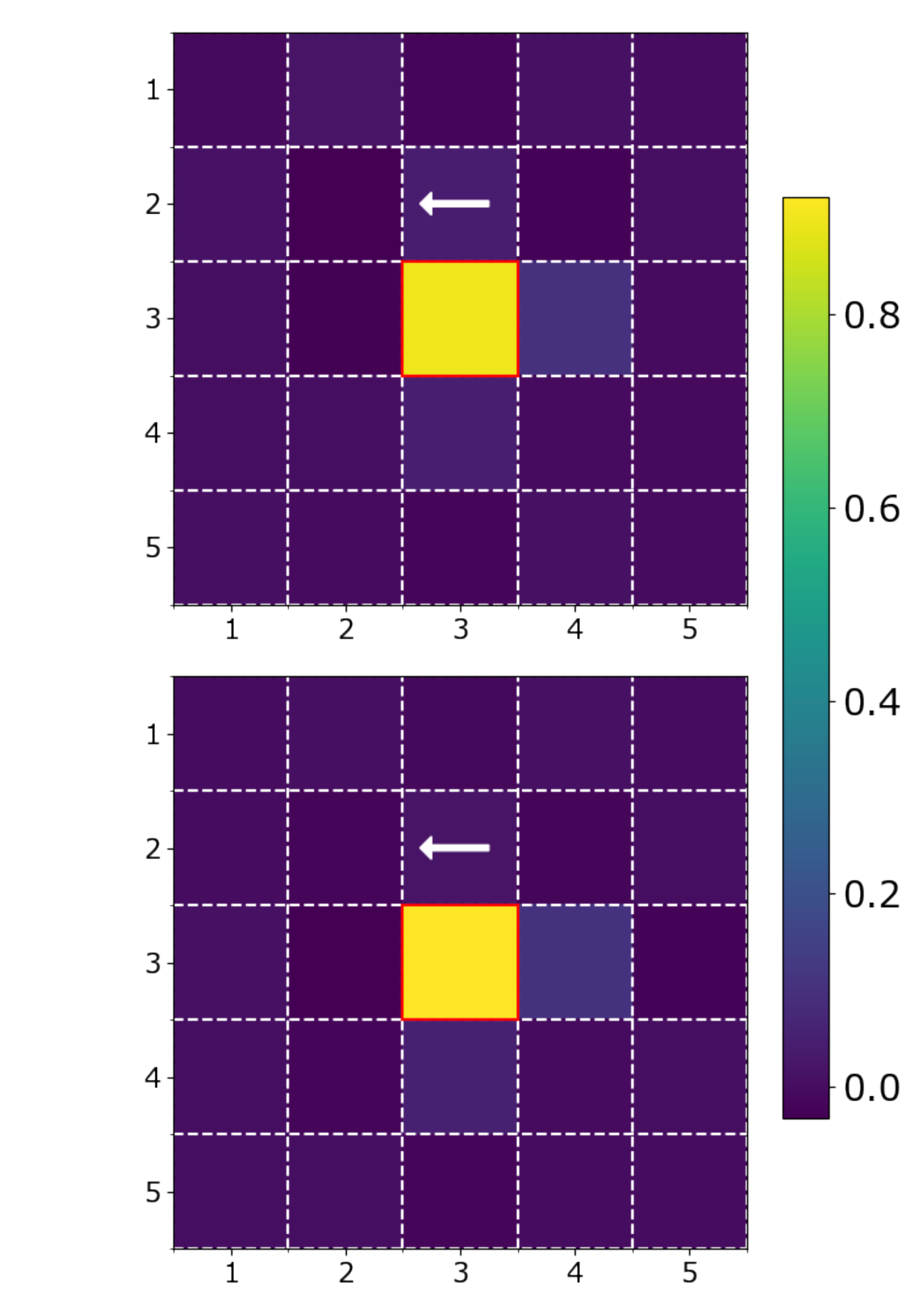}
\end{subfigure}
\begin{subfigure}{0.76\textwidth}
\includegraphics[height=6cm,trim={0 0cm 0 1cm},clip,width=\linewidth]{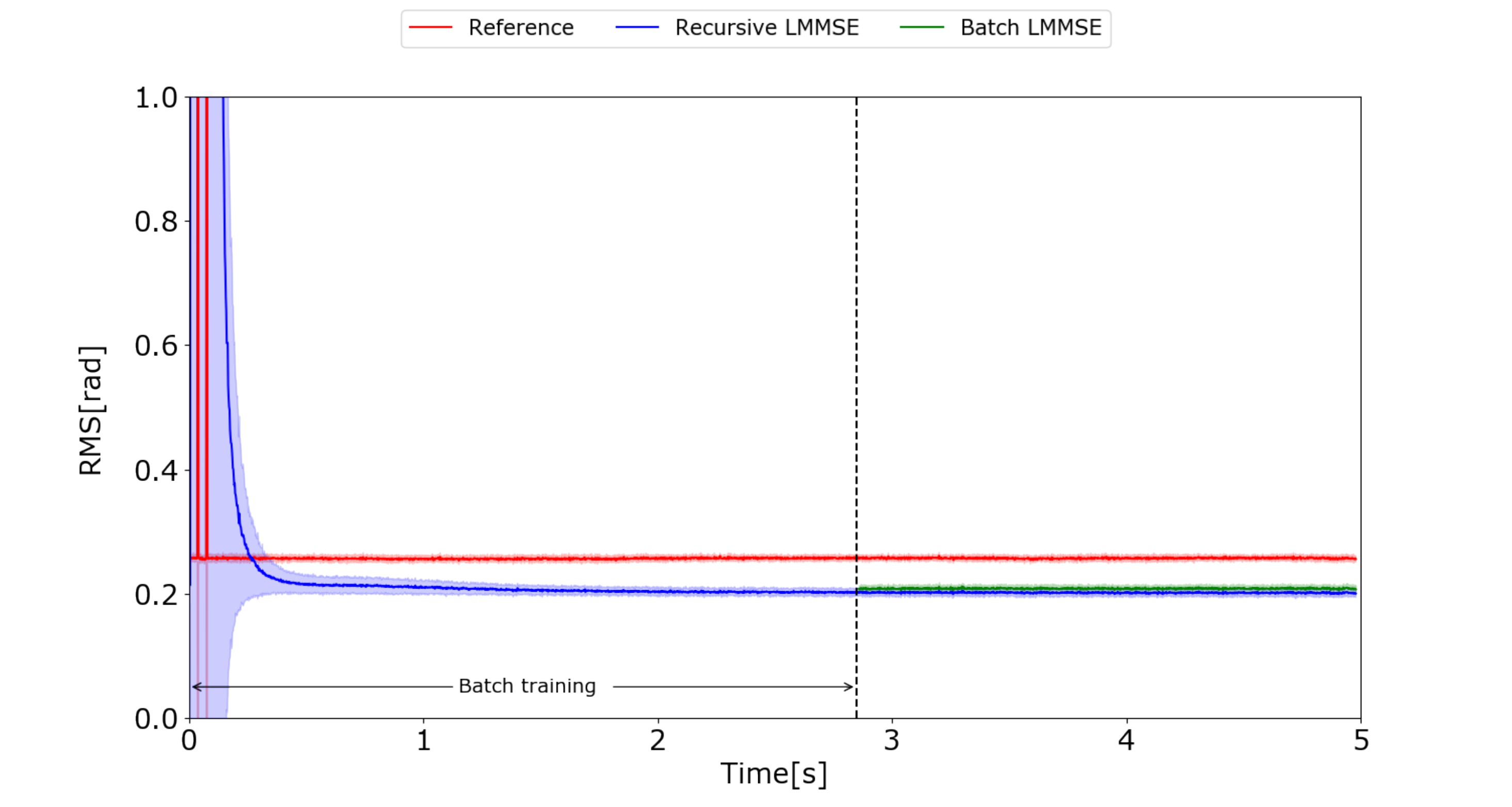}
\end{subfigure}\hspace*{\fill}
\caption{Left: The prediction coefficients for one phase point (outlined in red) for both the batch (top) and recursive (bottom) solutions. The arrow indicates the wind direction. Right: The mean residual wavefront error for multiple simulations as a function of time for our s1t1 classical solution (red line), batch LMMSE (green line), and recursive LMMSE (blue line).  The shaded regions indicate the 1-sigma levels. }
\label{fig:stationary_all}
\end{figure*}

\subsection{Stationary turbulence}\label{sec:results_stationary}
We limit the input data into the LMMSE algorithms by selecting a grid size of 5-by-5 phase points (regressors) of the most recent measurement to predict the central phase point 2 time steps into the future. For ease, we abbreviate our LMMSE with spatial order 5 and temporal order 1 to s5t1 (see Sect.~\ref{sec:prediction}). We then define our `s1t1' solution as our classical solution. For this case, the algorithm only sees the most recent measurement of one spatial point. This is a pure temporal solution with a history of 1 frame (the most recent measurement). We simply train it with a few frames and keep the solution static. Looking at the pure spatial LMMSE we see the prediction coefficients found by the LMMSE in a Frozen Flow case with a $r_0=0.13$ m (at 500 nm) and $v=10$ m/s, Fig.~\ref{fig:stationary_all}. Both the batch and recursive solutions show that the phase value at the yellow grid point has been identified as the largest contributor to the behavior of the central phase (outlined in red) at the next time step. Here the phase itself is the largest contributor. However, they both give a slight weighting ($\sim$0.1) to the neighboring point, indicating that the wind is shifting horizontally. The solution is not spread over many spatial points because the sub-apertures are so large and we are sampling the atmosphere at 1~kHz, with the phase screen traveling less than 0.5 sub-aperture per sample interval.

On the right in Fig.~\ref{fig:stationary_all} we plot the results for the stationary case.  We run 10 simulation experiments with the same input parameters (for all future runs as well). We calculate the root-mean-square (RMS) for the 10 residual phase screens at the same time instance. We shaded regions indicate the 1-sigma level.  We see that s5t1 performs better than our s1t1 for a single layer Frozen Flow atmosphere. This behavior holds true under a variety of   turbulence conditions (different values of $v$, $r_0$, and $L_0$).   This validates our LMMSE algorithms; under stationary conditions, we can use prediction to increase our performance compared to the static pure temporal solution s1t1.

\begin{figure}
\centering
\includegraphics[trim={0.5cm 0 1.5cm 0},clip, width=\linewidth]{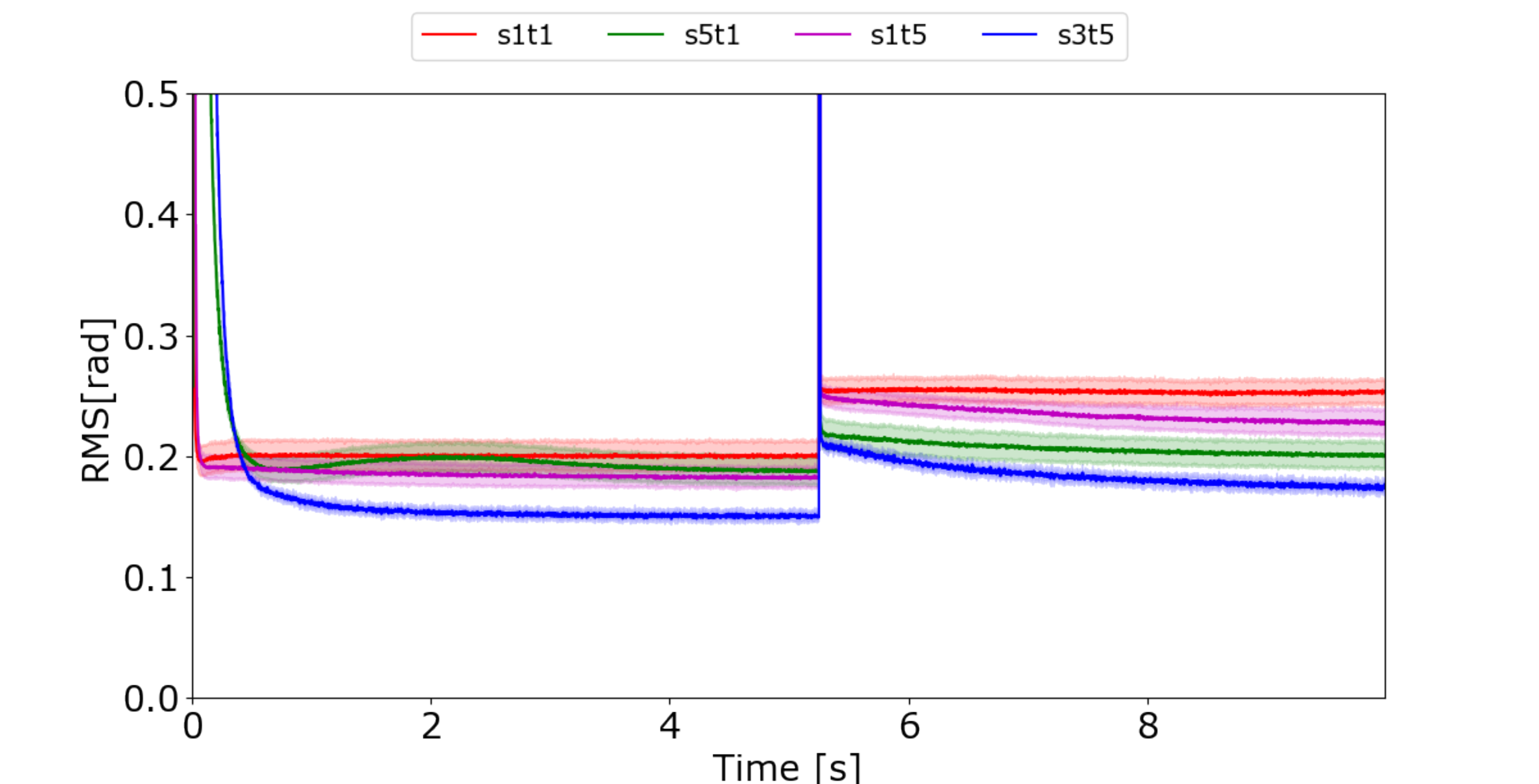}
\caption{Error and convergence of linear predictors with various regressor orders (top) before and after a jump in wind speed (bottom).  The shaded regions indicate the 1-sigma levels.  }
\label{fig:converge_plot}
\end{figure}

In the first 5 seconds of Fig.~\ref{fig:converge_plot} we show how different spatio-temporal predictors with different orders converge differently and are even able to achieve different final performance levels despite having converged fully for different amounts of spatio-temporal information. The amount of data in the solution controls how fast the solution fully converges as well as the residual variance. For higher order spatio-temporal solutions, the convergence rate is slower but the residual variance is smaller. It is important to note that the amount of input data increases much more for increases in spatial information as we radially increase the spatial information (as opposed to temporal information) as to not be biased towards a specific wind direction. The LMMSE is convenient in that it allows us to choose the amount of input data used for prediction to minimize computation time - either through an identification routine (i.e., single value decomposition while finding the batch solution or manually based on prior knowledge). We choose to limit the amount of information by knowledge of the physical constraints of the AO system such as the spatial sampling of the wavefront sensor and the temporal resolution as well as wind conditions.   For example, for our system we expect a maximum of 14 $m/s$ winds, therefore, our turbulence induced phase, in one time step, does not move more than one subaperture. Hence, we need one spatial subaperture minimum to capture most of the effects. We chose to increase this to two subapertures to provide edge information. However, to maintain symmetry as to not have preference to a wind direction, we land up with a 3-by-3 grid.  Not only is this attractive to reduce computation but for recursive methods, allows for the solution to converge faster.

%%%%%%%%%%%%%%%%%%%%%%%%%%%%%%%%%%%%%%%%%%%%%%%%%%%%%%%%%%%%%%%%%%%%%%%%%%%%%
\subsection{Effects due to time-varying wind speed} \label{sec:non_stationary}
With knowledge of how the LMMSE behaves under stationary conditions, we can study the effects of time varying atmospheric turbulence on our predictors. For this work we set the regressors to be a 3-by-3 spatial grid for 5 previous measurements for each phase point (s3t5). We focus our attention on the effect of a varying wind.

To start, we look at a very simple case for varying wind. Fig.~\ref{fig:converge_plot} shows the effect of an instantaneous wind speed jump of 7 m/s. First, we note that as the wind speed increases, the benefit of our predictor increases (i.e., the relative difference between the s3t5 solution and the classical solution). We can see that the recursive algorithm takes a long time (a few seconds) to converge back to its optimal solution after a change in wind speed.   

\begin{figure}
\centering
\includegraphics[trim={1cm 0 2cm 0},clip,width=\linewidth]{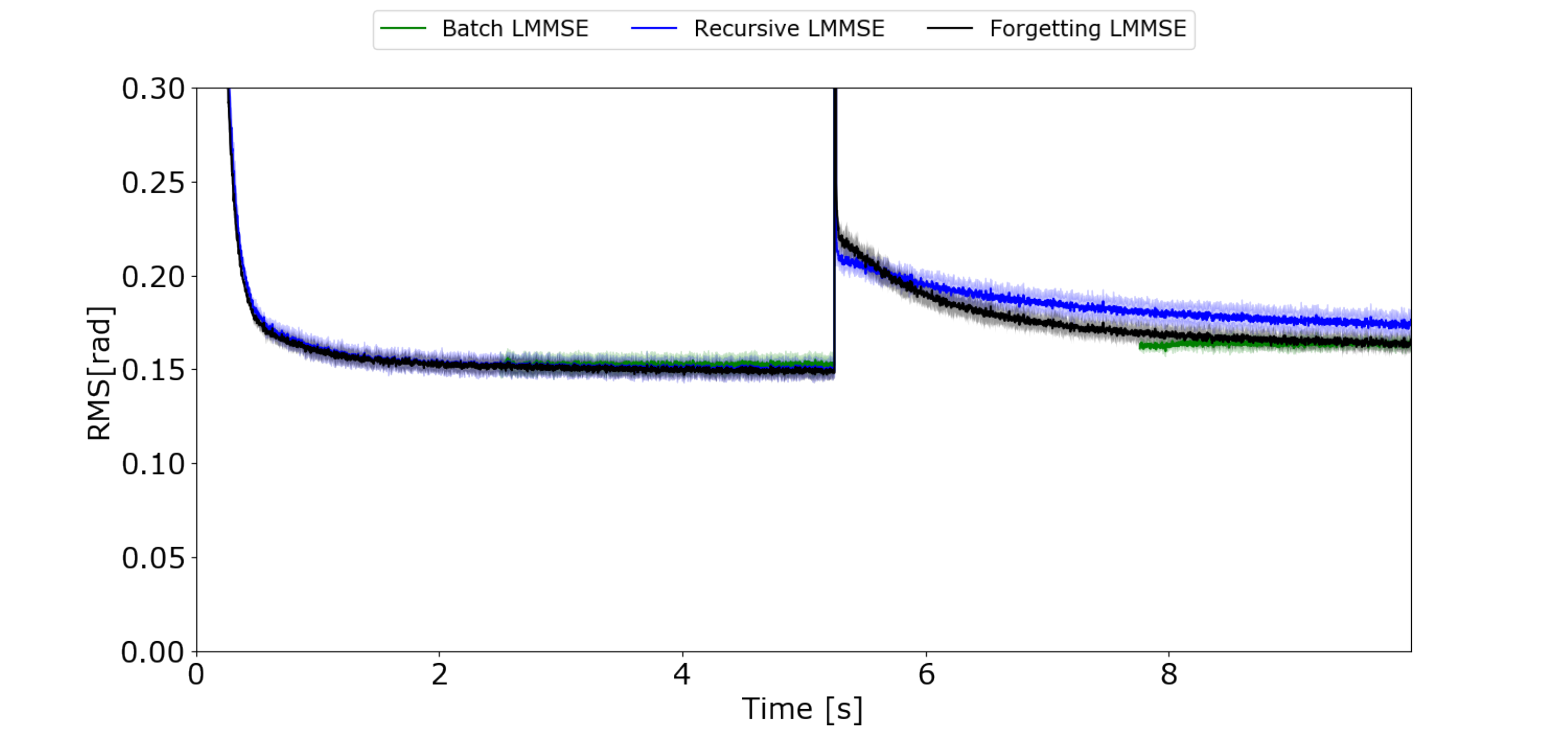}
\caption{ Wind jump of 7 m/s on a METIS scaled system showing the convergence of the recursive LMMSE (blue) and the forgetting LMMSE (black) as well as the resetting of the batch LMMSE (green) for s3t5.}
\label{fig:adaptive}
\end{figure}
\begin{figure*} % "[t!]" placement specifier just for this example
\begin{subfigure}{0.48\textwidth}
\includegraphics[trim={0 1.5cm 2cm 0},clip,width=\linewidth]{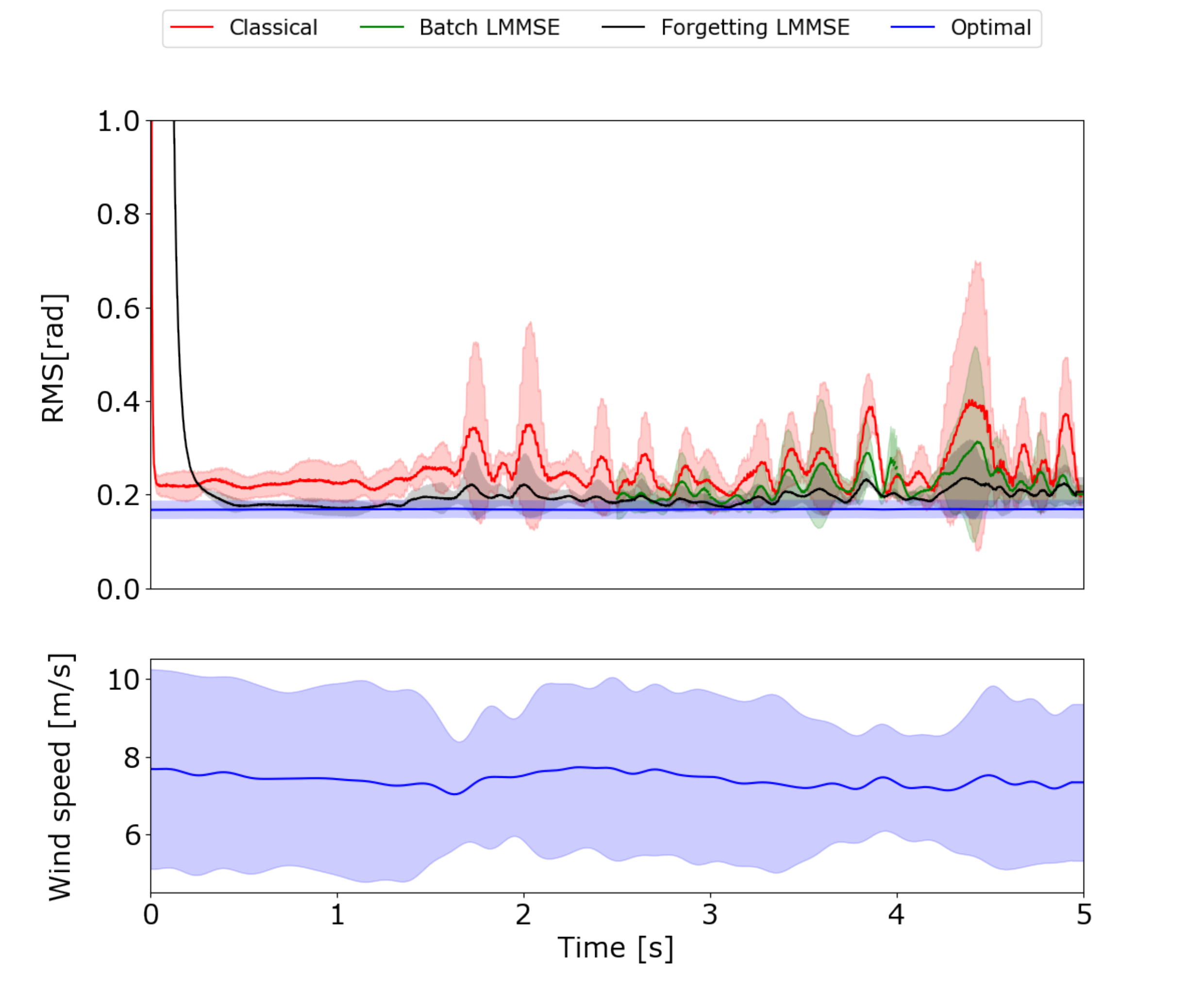}
 \caption{2007.06.15, $L_0 = 24$~m}\label{fig:a}
\end{subfigure}\hspace*{\fill}
\begin{subfigure}{0.48\textwidth}
\includegraphics[trim={2cm 1.5cm 0 0},clip,width=\linewidth]{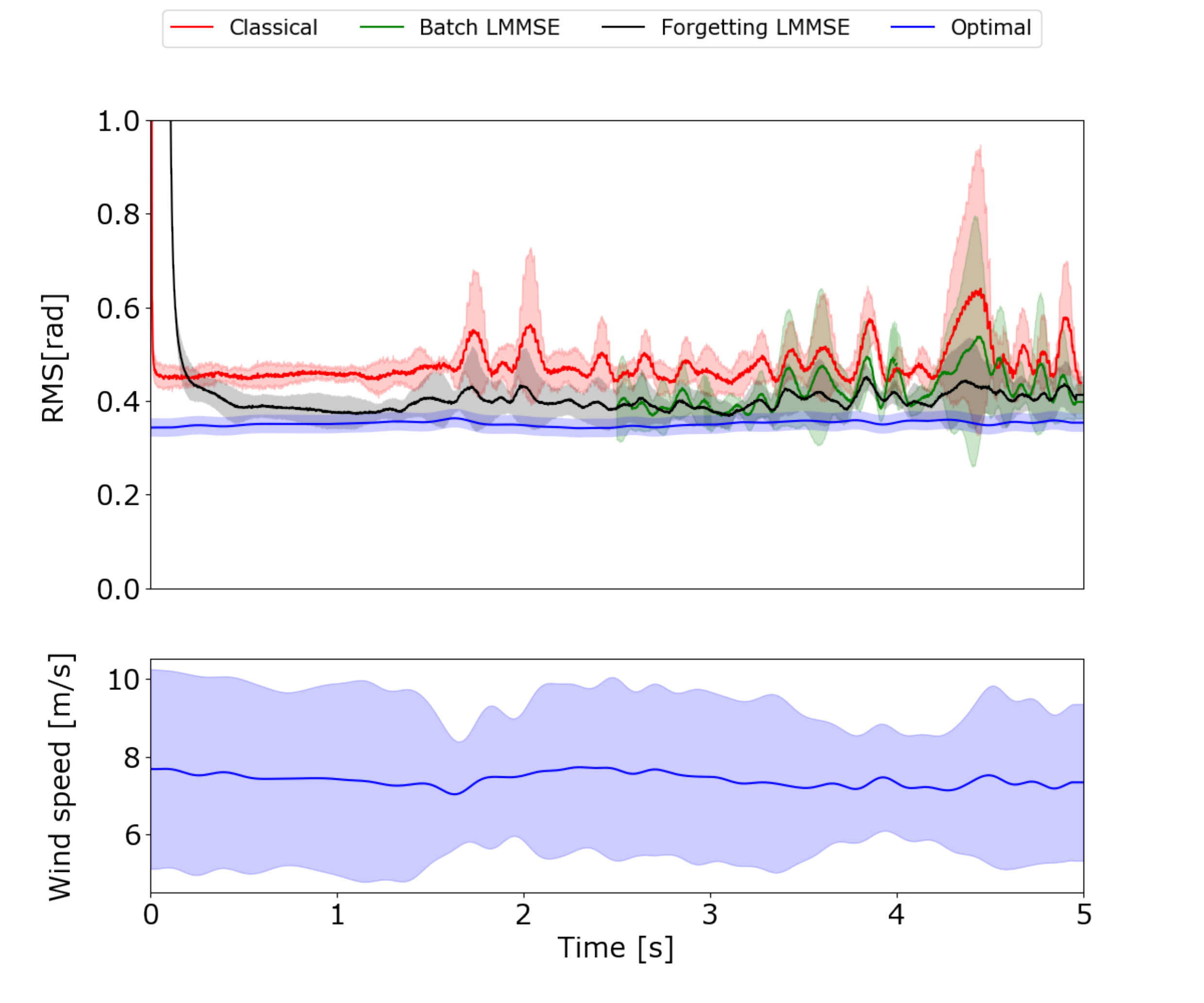}
\caption{2007.06.15, $L_0 = 80$~m}\label{fig:b}
\end{subfigure}

\medskip
\begin{subfigure}{0.48\textwidth}
\includegraphics[trim={0 1.5cm 2cm 2cm},clip,width=\linewidth]{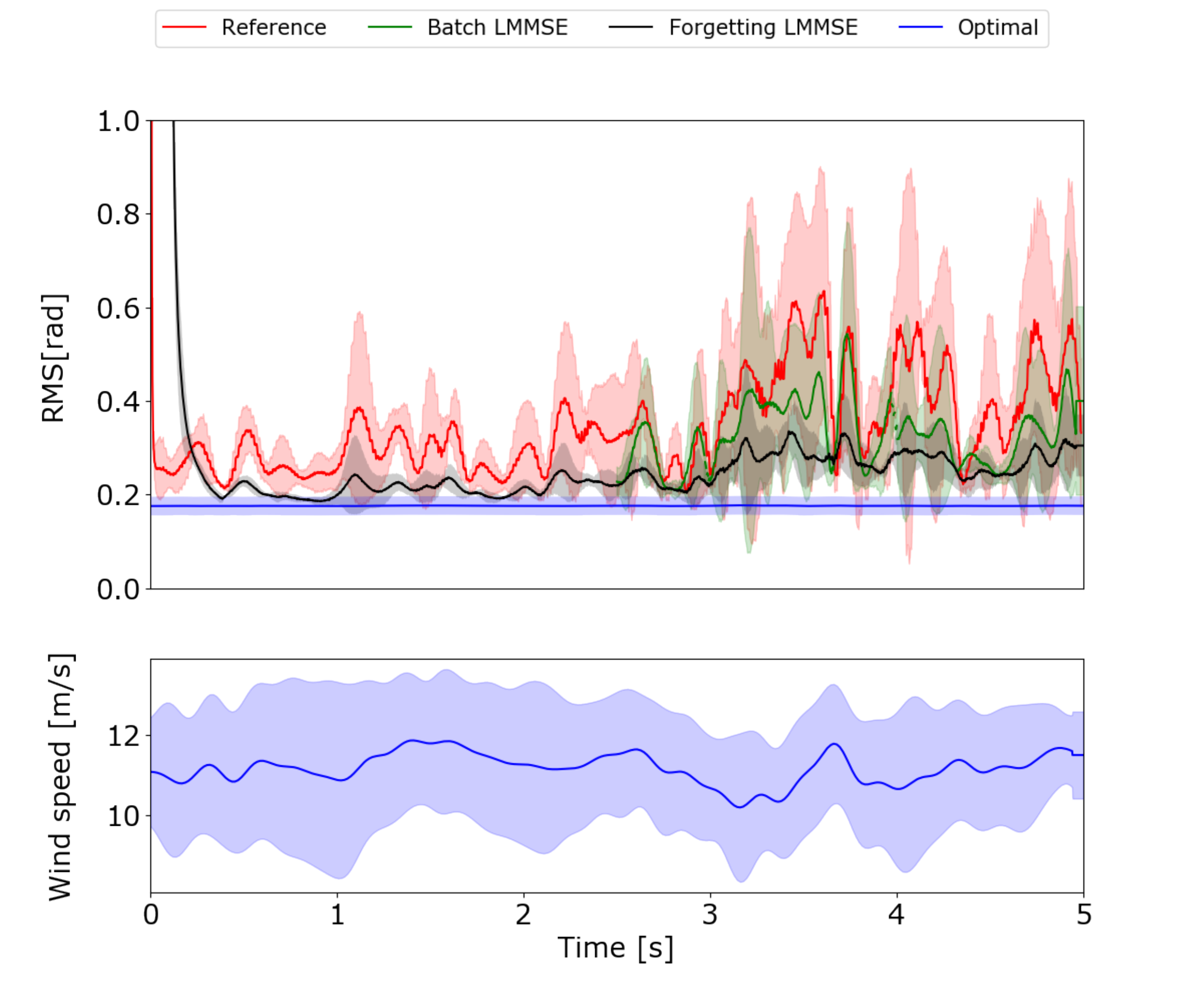}
\caption{2008.05.09, $L_0 = 24$~m} \label{fig:c}
\end{subfigure}\hspace*{\fill}
\begin{subfigure}{0.48\textwidth}
\includegraphics[trim={2cm 1.5cm 0 2cm},clip,width=\linewidth]{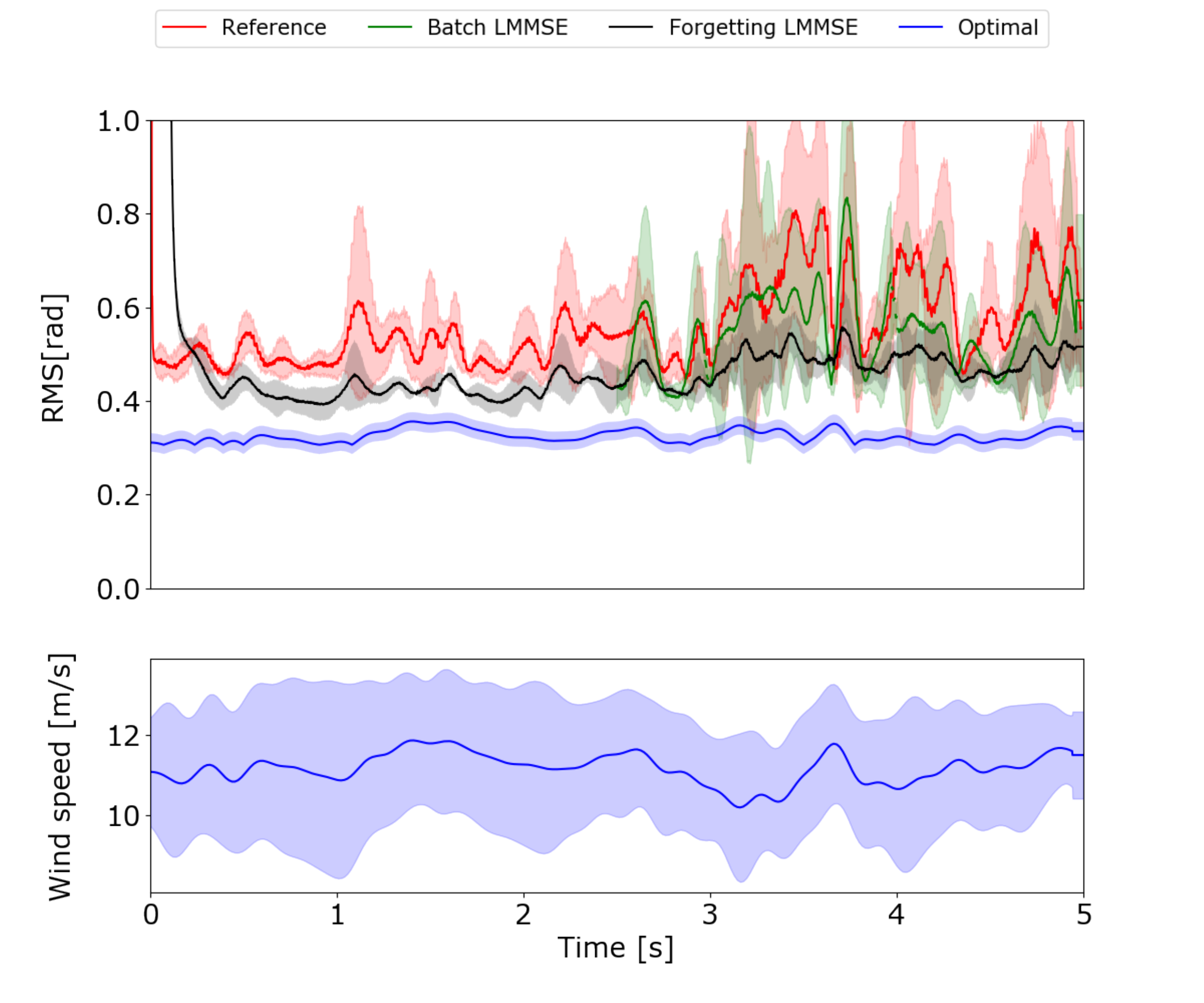}
\caption{2008.05.09, $L_0 = 80$~m}  \label{fig:d}
\end{subfigure}

\medskip
\begin{subfigure}{0.48\textwidth}
\includegraphics[trim={0 0 2cm 2cm},clip,width=\linewidth]{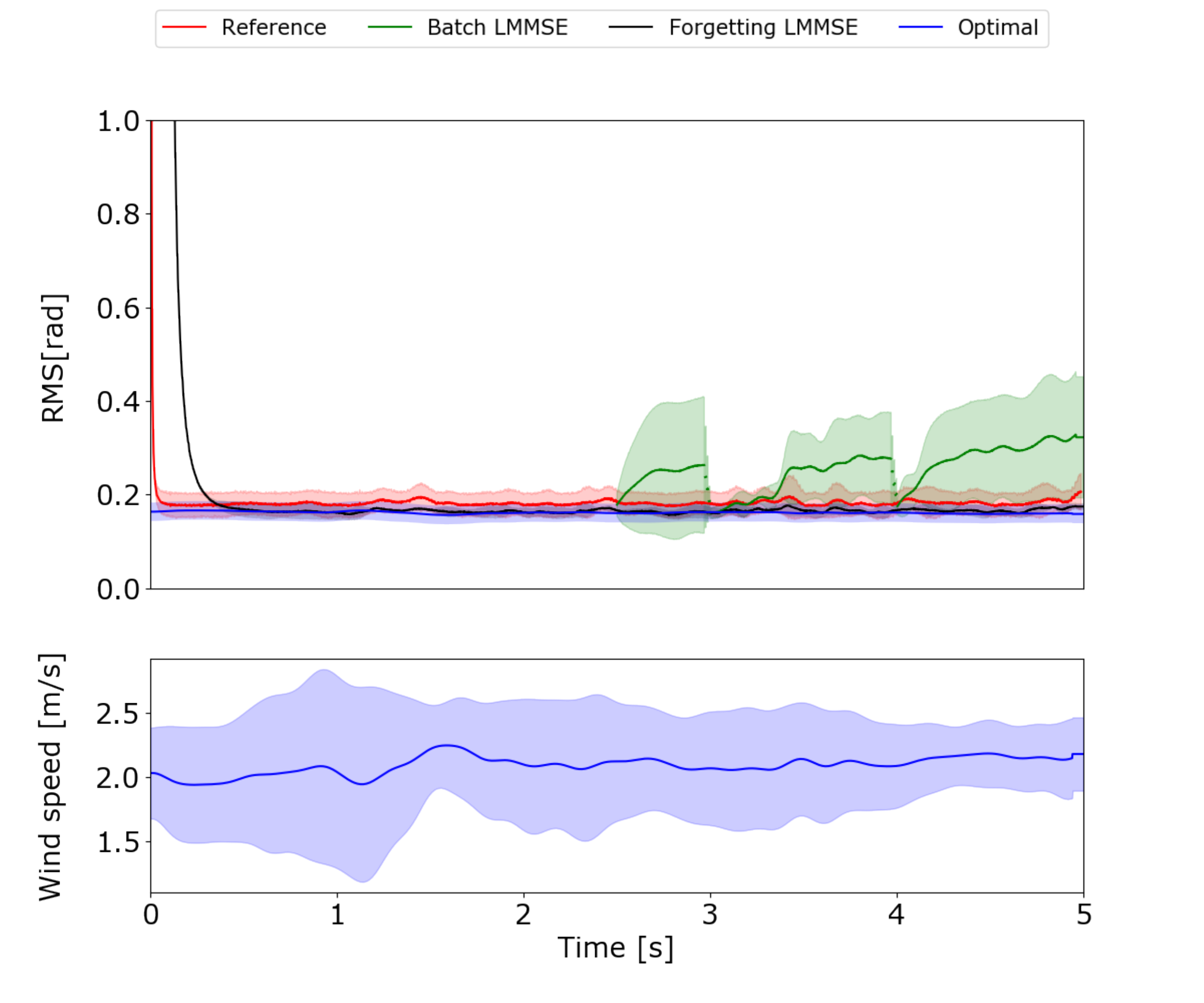}
\caption{2006.11.22, $L_0 = 24$~m} \label{fig:e}
\end{subfigure}\hspace*{\fill}
\begin{subfigure}{0.48\textwidth}
\includegraphics[trim={2cm 0 0 2cm},clip,width=\linewidth]{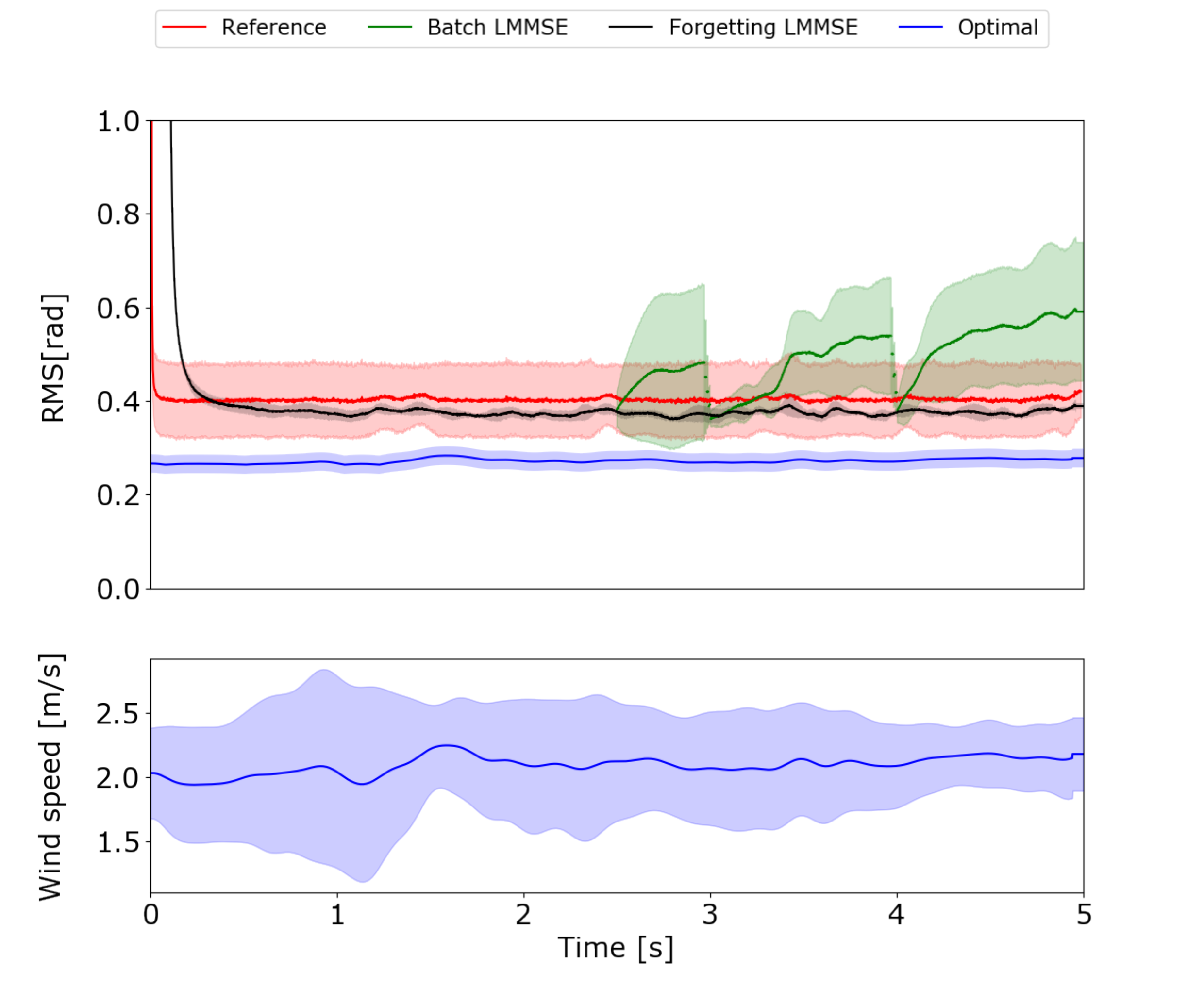}
\caption{2006.11.22, $L_0 = 80$~m} \label{fig:f}
\end{subfigure}
\caption{The average response for our predictors (upper panel) for three different wind conditions corresponding to Tab. 1 (lower panel), with $L_0~=~24$~m (left column) and $L_0~=~80$~m (right column).  We plot the s1t1 predictor (red), the s3t5 resetting batch LMMSE (green), the s3t5 forgetting LMMSE (black), and the s3t5 ideal LMMSE (blue). We first allow the batch to train on 2.5 s of data and then reset it every 1s.  } \label{fig:1}
\end{figure*}

In Fig.~\ref{fig:adaptive} we compare the three different types of LMMSE algorithms for our s3t5 solution under a wind speed jump. The first being the non-adaptive solution: the recursive LMMSE predictor. The recursive LMMSE predictor  updates the covariance estimate when new data becomes available and in turn the prediction coefficients (Eqs. ~\ref{eq:cov1} to ~\ref{eq:a_update}). The estimation makes use of all previous data. Therefore, it will not be able to forget the data before the jump and in this example it takes longer to converge after the jump than its initial convergence.  By knowing the moment when the wind speed changes we are able to re-calibrate our batch LMMSE predictor to minimize the loss in performance and provide the optimal solution- this is one type of adaptive solution. Finally, we show our second adaptive algorithm (and third LMMSE algorithm), the forgetting LMMSE predictor. We can see that after the wind jump the forgetting converges slightly faster than the recursive due to the forgetting factor.

\subsubsection{Wind speed fluctuations}\label{sec:wind_fluc}
We now look at the behavior for wind speed fluctuations, Fig.~\ref{fig:1}. We do this for two different $L_0$ values: one smaller than the telescope diameter (24 m) and one larger than the telescope diameter (80 m). We plot our adaptive LMMSE solutions: the forgetting LMMSE and the resetting batch LMMSE. As in reality, the batch has no knowledge of the wind speed behavior and therefore it is arbitrarily reset. For the classical solution, we once again plot the s1t1 solution. Finally, we plot our ideal s3t5 predictor. The ideal predictor is the s3t5 solution under infinite tracking speed and mimicking an instantaneous stationary case. We pre-determine the s3t5 predictor for each potential wind speed assuming stationary conditions and then use a look-up table to find our ideal predictor performance for Fig.~\ref{fig:1}. This provides us with a lower bound of a s3t5 predictor's performance. We run the simulation for multiple cases of varying wind on the same night and average the results. This is done for the three different nights shown in Sect.~\ref{sec:wind}. We plot the mean wind speed and 1-$\sigma$ (standard deviation) levels (shaded region) for the wind speed as well as the different predictors.

\section{Analysis}\label{sec:discussion}
 
\begin{table*}
\centering
\begin{tabular}{ccccccc}
\hline
Date & 2007.06.15 (June) & & 2008.05.09 (May)& & 2006.11.22 (Nov)  &   \\ 
\hline\hline
Outerscale & 24 m & 80 m & 24 m & 80 m & 24 m & 80 m \\
\hline
Classical & 1.42&1.40 &1.96 & 1.59 & 1.19 &1.49\\ 
Batch LMMSE & 1.21 &1.19&1.83 &1.55 & 1.59 & 1.77 \\
Forgetting LMMSE &1.08 & 1.13 & 1.47 &1.35 &1.04 &1.36 \\
\hline
\end{tabular}
\caption{Summary of Fig~\ref{fig:1} showing the ratio of the mean rms value of an algorithm to the optimal LMMSE.}
\label{table:2}
\end{table*}

 We summarize the results of  Sect.~
\ref{sec:non_stationary} and Fig.~\ref{fig:1} in Tab.~\ref{table:2}. We arrive at the following conclusions:
\begin{enumerate}
\item All our LMMSE approaches lose performance when compared to the ideal predictor under varying wind (the ratio between the batch and forgetting LMMSE is greater than 1).
\item The forgetting implementation does better than the resetting batch implementation (the forgetting LMMSE ratio is smaller than the batch LMMSE ratio).
\item The difference between s3t5 and the classical solution becomes smaller under varying wind.
\item The optimal solution has a much smaller variance compared to the other algorithms.
\end{enumerate}
We conclude that a different predictive control approach (that can either track faster or is less sensitive to time-varying fluctuations) is needed to predict atmospheric turbulence in the presence of time-varying turbulence as it is noticeably different than the stationary case. 
 
More specifically, from the plots in Fig.~\ref{fig:1} we conclude that the benefits of our LMMSE predictors are  modest . From the optimal solution, we not only lose in residual mean wavefront error (see Tab.~\ref{table:2}) but also have much larger variances for our LMMSE. Ideally, a predictor would also provide a stable correction with a small variance such as the optimal solution.  

  In all three wind cases, we see the forgetting LMMSE predictor is able to do fairly well for the first couple of seconds for $L_0=24$, reaching the ideal solution. However, the algorithm is unable to handle the time-varying fluctuations as seen by the increase in variance compared to the optimal solution that has unlimited tracking capabilities. We attribute this to the continuously changing wind. As we see in Fig.~\ref{fig:adaptive}, the forgetting algorithm needs between 0.5~s to 1~s to recover from a large jump, in part due to the number of regressors, and in part due to the magnitude of the change in wind speed. The forgetting LMMSE predictor performs better for slowly evolving changes than fast changes. From Tab.~\ref{table:2} we see that the forgetting LMMSE does better than the resetting batch predictor and the classical approach in every case.   

 %%%%%%%%%%%%%%%%%%%%%%%%%%%%%%%%%%%%%%%%%%%%
\subsection{Further Discussion}\label{sec:further_dis}
From the stationary tests we can conclude that the LMMSE algorithms with higher order spatio-temporal solutions have a benefit compared to a pure temporal solution making use of only the most recent measurement as we would expect from using Frozen Flow to generate our atmospheric turbulence phase. This agrees with the literature that prediction (whether model-based or data-driven) can provide a reduction in the servo-lag error and the overall residual wavefront error. 

\begin{figure}[ht!]

\includegraphics[trim={1cm 0 2cm 0},clip,width=\linewidth]{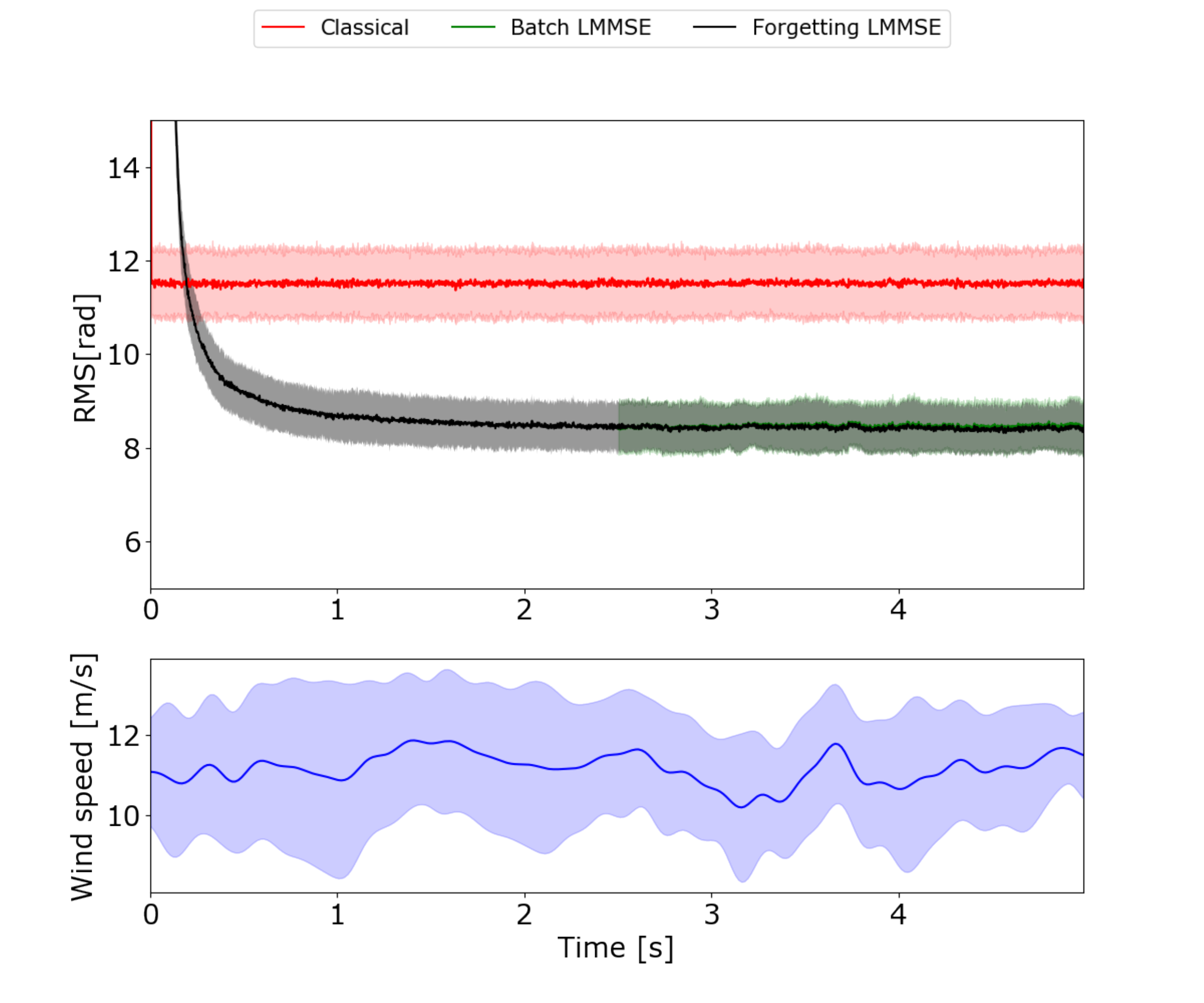}
\caption{Prediction performance for the case of the noise being 0.5 times the turbulence variance.}
\label{fig:noise}
\end{figure}

The wind fluctuations reported by the TMT site testing group adversely impact the basic predictor - the LMMSE - except for the specific case of slowly moving wind speed. Wind speeds can fluctuate by over 1~m/s on time scales less than 1~s. Under time-varying conditions we would expect only a minor performance improvement for a single conjugate AO system with a low order LMMSE predictor compared to no prediction when the system is dominated by the servo-lag error. It is important, however, to note that due to the measurement height of 7~m these measurements might not reflect what is actually seen by an AO system as the dome and other buildings will influence the low layer winds. These fluctuations might also be extreme when compared to other astronomical sites.  Finally, by limiting our analysis to the ground layer (due to available data), we might be overestimating the strength and timescales of the wind fluctuations and therefore the effect on prediction. 

Assessing our prediction in the presence of our wind fluctuations we quickly see why on-sky test campaigns for predictive AO control are more challenging than common laboratory tests. A single wind jump step already provides a challenge for the LMMSE, moving it away from the ideal predictor solution (Fig.~\ref{fig:converge_plot}). When we extend our simulation to real wind data, the LMMSE has a harder time tracking the changes (Fig. ~\ref{fig:1}). Therefore, time-variant behavior needs to be included both in simulation and in the laboratory setting as well as predictor design

In the above work we have looked only at the specific case where the temporal delay is a main limiter in performance and hence the use of a LMMSE as a predictor over the time delay only. However, we can also benefit from using these algorithms when we have other dominating error terms. For an increase in a factor of 25 in measurement noise we plot the performance of our high order LMMSE predictors in Fig.~\ref{fig:noise}. We are able to perform better than the classical solution for varying winds. 
 
%%%%%%%%%%%%%%%%%%%%%%%%%%%%%%%%%%%%%%%%%%%%
\section{Conclusions}\label{sec:conclusion}
In this paper, we model dynamic wind behavior, based on the TMT site testing data. We look at different implementations of a low order LMMSE predictor to predict atmospheric phase fluctuations over a time delay of two frames. For the stationary case, both the batch and recursive implementations perform better than a pure temporal predictor; i.e., that the current measurement is the best prediction. Under time varying turbulence, however, we loose performance due to wind speed fluctuations. For slower wind speeds and $L_0$ smaller than the telescope diameter, we are able to approach ideal performance with our LMMSE. The classical solution is also able to perform well under these conditions.   As the wind speed increases, we see a larger difference between the LMMSE and the ideal solution.  For our given spatial and temporal sampling there are conditions (smaller $L_0$ and wind speeds approximately 8 m/s) in which we do reach an optimal predictor solution.  

This work demonstrates that for an AO system the process of selecting and designing a predictive control framework for time varying wavefront phase fluctuations is much different than for the stationary case. In ongoing research, we are aiming at more suitable predictive control schemes that will be better able to deal with the statistical variability of atmospheric turbulence, especially for cases of high wind speed variance.

\section{Acknowledgments}
The authors would like to thank TMT site testing group, specifically Warren Skidmore and Tony Travouillon, for providing us with the wind speed data. The  authors  would also like  to  thank  Leiden  University, NOVA,  METIS  consortium,  and  TNO  for  funding  this research
%%%%%%%%%% If using BibTeX:
\bibliography{sample}

\end{document}